# MyoMapNet: Accelerated Modified Look-Locker Inversion Recovery Myocardial $T_1$ Mapping via Neural Networks

Hossam El-Rewaidy[1,2], Rui Guo[1], Amanda Paskavitz[1], Tuyen Yankama[1], Long Ngo[1], Bjoern Menze[2,3], Reza Nezafat[1]

[1]Department of Medicine, Cardiovascular Division, Beth Israel Deaconess Medical Center and Harvard Medical School, Boston, Massachusetts, USA

[2]Department of Computer Science, Technical University of Munich, Munich, Germany

[3]Department for Quantitative Biomedicine, University of Zurich, Switzerland

---

Running title: Fast MOLLI $T_1$ Mapping

Word Count: 4700

**Address for correspondence**: Reza Nezafat, PhD

Department of Medicine, Beth Israel Deaconess Medical Center,

330 Brookline Avenue, Boston, MA 02215

Phone: +1- 617-667-1747; Fax: +1- 617-975-5480

E-mail: rnezafat@bidmc.harvard.edu

# Abstract

**Purpose**: To develop and evaluate MyoMapNet, a rapid myocardial $T_1$ mapping approach that uses neural networks (NN) to estimate voxel-wise myocardial $T_1$ and extracellular (ECV) from $T_1$-weighted images collected after a single inversion pulse over 4-5 heartbeats.

**Method:** MyoMapNet utilizes a simple fully-connected NN to estimate $T_1$ values from 5 (native) or 4 (post-contrast) $T_1$-weighted images. Native MOLLI-5(3)3 $T_1$ was collected in 717 subjects (386 males, 55±16.5 years) and post-contrast MOLLI-4(1)3(1)2 in 535 subjects (232 male, 56.5±15 years). The dataset was divided into training (80%) and testing (20%), where 20% of the training set was used to optimize MyoMapNet architecture (size and loss functions). We used MyoMapNet to estimate $T_1$ and ECV maps with the first 5 (native) or 4 (post-contrast) $T_1$-weighted images from the corresponding MOLLI sequence compared to the conventional and an abbreviated MOLLI using similar number of $T_1$-weighted images with 3-parameter curve-fitting.

**Results:** In our preliminary optimizaiton step, we determined that a 5-layers NN trained using mean-absolute-error loss yields lower estimation errors and was used subsequently in independent testing study. The myocardial $T_1$ by MyoMapNet was similar to MOLLI (1200±45ms vs. 1199±46ms; P=0.3 for native $T_1$, and 27.3±3.5% vs. 27.1±4%; P=0.4 for ECV). MyoMapNet had significantly smaller errors in $T_1$ estimations compared to abbreviated-MOLLI (1±17ms vs. 31±34ms, P<0.01 for in native $T_1$, and 0.1±1.3% vs. 1.9±2.5%, P<0.01 for ECV). The duration of T1 estimation was approximately 2 ms per slice using MyoMapNet.

**Conclusion:** MyoMapNet $T_1$ mapping enables myocardial $T_1$ quantification in 4-5 heartbeats with near-instantaneous map estimation time with similar accuracy and precision as MOLLI.

## Introduction

Cardiovascular magnetic resonance (CMR) myocardial $T_1$ and extracellular volume (ECV) mapping allow for non-invasive quantification of interstitial diffuse fibrosis (1). Over the past decade, there have been significant advances in CMR pulse sequences for myocardial $T_1$ mapping (2–11). These sequences are based on the application of magnetization preparation pulses (such as inversion (2,4,6), saturation (3,5), or a combination of both (8)) for the collection of a series of $T_1$ weighted images. Subsequently, tissue relaxation times are estimated using a two or three-parameter fitting model (2–12). For each of these sequences, there are different trade-offs in accuracy, precision, coverage (2D vs. 3D), and respiratory motion (free breathing vs breath-holding) (11,13–23). Alternatively, recent approaches, such as combined $T_1$ and $T_2$ mapping (21,24–27), MR fingerprinting (28,29), or multitasking (30), have been introduced to simultaneously measure different tissue parameters.

Among different myocardial $T_1$ mapping sequences, Modified Look-Locker (LL) inversion recovery (MOLLI) (2), acquired within a single breath-hold, is the most widely used sequence for myocardial $T_1$ mapping. This sequence consists of three sets of LL experiments with 3 heartbeats (or 3 seconds) in between for magnetization recovery. This acquisition is commonly referred to as the 3(3)3(3)5 MOLLI scheme, which indicates that 3, 3, and 5 images are acquired in three LL experiments with 3 rest heartbeats (i.e., 3RR interval) between each LL experiment. MOLLI-5(3)3 and MOLLI-4(1)3(1)2 were implemented to improve acquisition efficiency and precision (13,31). A variation of these sequences in which the rest period between acquisition utilizes a fixed time (1-3 second) instead of heartbeats was also implemented (32). These modifications reduce the duration of the original MOLLI from approximately 17 heartbeats to 11 heartbeats (13). In this approach, $T_1$-weighted images are acquired at multiple points throughout the recovery curve, and a pixel-wise curve fitting algorithm using a 3-parameter model is applied to estimate the $T_1$ values at each pixel. Recently, the inversion group (IG) fitting has also been proposed for MOLLI with shorter waiting periods between LL experiments, albeit with a penalty on precision (33–35). For shortened MOLLI (ShMOLLI) with a 5(1)1(1)1 scheme (4), a conditional fitting algorithm is utilized to discard the latter measurements for long $T_1$ at high heart-rates.

Alternatives to standard curve-fitting techniques in parametric mapping include dictionary-based reconstruction, simulated signal recovery, or machine learning (10,36–39). Shao et al. used a deep learning model for rapid and accurate calculation of myocardial $T_1$ and $T_2$ in Bloch equation simulation with slice profile correction (36). Similarly, Hamilton et. al. used deep learning to rapidly reconstruct $T_1$ and $T_2$ maps from CMR fingerprinting images (37). To reduce motion artifacts and minimize the acquisition window in $T_1$ mapping during the cardiac cycle, a convolutional neural network (CNN) model was used to reconstruct highly accelerated $T_1$ weighted images in slice-interleaved $T_1$ mapping sequences with radial sampling (38). Deep learning neural networks were also recently used to combine saturation recovery and inversion data to improve $T_1$ mapping precision (40). These studies have demonstrated the potential for machine learning to improve myocardial tissue characterization by increasing precision and reconstruction speed, or decreasing motion sensitivity or other confounders of $T_1$ mapping.

In this study, we sought to develop and evaluate a fully connected neural network (MyoMapNet)-based rapid myocardial $T_1$ mapping approach using a single LL experiment with a scan duration of 4-5 heartbeats and near-instantaneous (~2ms) reconstruction time. The proposed MyoMapNet was trained and validated using numerically simulated data and in-vivo images; MyoMapNet's performance was compared to MOLLI with a 3-parameter fitting model.

# Methods

## MyoMapNet $T_1$ Mapping

The data acquisition for MyoMapNet is very similar to MOLLI sequence and is based on the collection of 5 (native) or 4 (post-contrast) $T_1$ weighted images after a single inversion pulse. **Figure 1** shows the pulse sequence and magnetization evolution of MOLLI-5(3)3 for native $T_1$ mapping, composed of two LL experiments to acquire eight $T_1$ weighted images over 11 heartbeats. In MyoMapNet, we use the five images sampled after the first inversion pulse for native MOLLI-5(3)3 $T_1$ measurement. For post-contrast $T_1$ mapping, we will use only the first four images collected using MOLLI-4(1)3(1)2 $T_1$ (**Supplementary Figure S1**). In contrast to MOLLI, where a 3-parameter fit model is used to estimate native or post-contrast $T_1$ mapping, in MyoMapNet, we propose to use a neural network to estimate $T_1$ values from only 5 (native) or 4 (post-contrast) $T_1$ images, thereby reducing the scan time to only 5 heartbeats for native and 4 heartbeats for post-contrast scans.

### Numerical simulation

MOLLI-5(3)3 simulations were performed to generate the training and testing signal-intensity time courses. A single-shot readout for two sequences was simulated using Bloch equation with the following parameters: balanced Steady-State Free Precession (bSSFP) acquisition with 5 ramp-up pulses, TR/flip angle =2.5 ms/35º, 80 phase-encode lines, acquisition window = 200 ms, and linear phase ordering. The time between the inversion and the center of k-space was 120 ms and 200 ms for two LL experiments of MOLLI-5(3)3. Simulated $T_1$ ranged from 400 ms to 1800 ms with an increment of 0.1 ms, $T_2$ of 42 ms, and a fixed heart rate of 60 bpm. Subsequently, different Gaussian noise levels were added to the simulated signals for SNRs of 20, 40, and 100. This process was repeated 5 times for each SNR to generate a total of 80,000 signal time-courses, each with 8 time-points, to serve as data augmentation. A 3-parameter fitting model was used to calculate $T_1$ of each signal time course:

$$s(TI) = A - Be^{-TI/T_1^*} \quad [1]$$

where TI represents the inversion time, and A, B, and $T_1^*$ were three unknowns. $T_1$ was determined from the resulting A, B, and $T_1^*$ by applying the equation: $T_1$=$T_1^*$(B/A-1).

## In-vivo data

The study protocol was approved by the Institutional Review Board and written consent was waived. $T_1$ mapping data was collected as part of the clinical exam in all participants, and we retrospectively extracted $T_1$ mapping data. Patient data were handled in compliance with the Health Insurance Portability and Accountability Act. Imaging was performed using a 3T scanner (MAGNETOM Vida, Siemens Healthineers, Erlangen, Germany) equipped with body and spine phased-array coils.

$T_1$ mapping using MOLLI was collected in 717 subjects (386 males, 55±16.5 years) with a subset of 535 subjects (232 males, 56.5±15 years) having both native and post-contrast $T_1$ maps. All data were extracted retrospectively from patients who were referred for a clinical CMR exam for various cardiovascular indications from October 2018 to March 2020.

MOLLI-5(3)3 and MOLLI-4(1)3(1)2 were performed using the following parameters: bSSFP with 5 ramp-up excitations, FOV = 360×320 mm$^2$, voxel size = 1.7×1.7×8 mm$^3$, TR/TE = 2.5 ms/1.03 ms, flip angle = 35°, linear phase-encoding ordering, partial Fourier factor = 7/8, GRAPPA acceleration factor of 2 with 24 reference lines, acquisition window ~218 ms during diastole, slice gap 12 mm. Three slices in the short-axis view (SAX) were collected in three separate breath-holds with 10 second rest periods between breath-holds. Adiabatic tan/tanh inversion pulse was used. Both MOLLI-5(3)3 and MOLLI-4(1)3(1)2 used a minimum inversion time (TI) of 100 ms with a TI increment of 80 ms. For post-contrast measurement, MOLLI-4(1)3(1)2 data was collected 15-20 min after injection of Gd-DTPA at 0.1 mmol/kg (Gadavist, Bayer Pharma AG, Berlin, Germany). MOLLI-5(3)3 and MOLLI-4(1)3(1)2 $T_1$ were calculated using a 3-parameter curve fitting model (Equation 1).

## $T_1$ Estimation in MyoMapNet

MyoMapNet is a fully-connected neural network that estimates $T_1$ maps from input $T_1$-weighted images. MyoMapNet input layer has $2N_t$ nodes for $N_t$ $T_1$ weighted signals and their corresponding inversion times, where $N_t$=5 in the native $T_1$ network, and $N_t$=4 in the post-contrast $T_1$ network. MyoMapNet is consisted of $M_l$ layers each of $(200 \times S)$ nodes followed by $M_l$ layers of $(100 \times S)$ nodes each, and one layer of $(50 \times S)$ nodes; where $M_l$ controls the depth (total number of layers) of MyoMapNet, and $S$ is a scaling factor to control the number of trainable parameters in each layer. For example, **Figure 2** shows the MyoMapNet architecture at $M_l = 2$ and $S = 2$, where five hidden layers of 400, 400, 200, 200, 100 nodes in each layer were utilized. A leaky rectified linear unit activation function was applied after each hidden layer. MyoMapNet parameters were empirically optimized to balance between accurate $T_1$ mapping estimation and avoiding the risk of overfitting. To determine the optimal network depth and size, MyoMapNet performance was validated at $M_l = 1, 2, and$ 3 and $S = 1, 2,$ and 3. The number of parameters at different depths and scales of MyoMapNet is reported in **Table S1**.

MyoMapNet was implemented in Python using the PyTorch library version 0.41. Training and validation were performed on an NVIDIA DGX-1 system equipped with 8T V100 graphics processing units (GPUs; each with 32 GB memory and 5120 cores), central processing unit (CPU) of 88 core: Intel Xeon 2.20 GHz each, and 504 GB RAM. Only 2 GPUs were considered for training and testing the proposed MyoMapNet.

## MyoMapNet Training

For simulation experiments, a set of 64,000 simulated signals (80% of the whole dataset) was randomly selected at each noise level (i.e. SNR of 100, 40, and 20). Each of these datasets was used to train the MyoMapNet separately. For in-vivo studies, training datasets of 573 patients (1719 native $T_1$ maps) and 428 patients (1281 post-contrast $T_1$ maps) were randomly selected for the native and post-contrast map reconstructions, respectively. A detailed description of data splitting is included in **Supplementary Figure S2**. Corrupted native $T_1$ maps (249 maps) due to image artifacts were excluded from the training dataset. MOLLI-5(3)3 and MOLLI-4(1)3(1)2 $T_1$ maps were estimated using 8 (native) or 9 (post-contrast) $T_1$-weighted images. We performed all the map reconstruction offline without motion correction. Two loss functions were studied to minimize the error between the estimated and reference MOLLI $T_1$ values:

a) The mean-absolute error loss function (MAE) $= \frac{1}{n}\sum_{i=1}^{n}\left|estimated_T_{1_i} - MOLLI_T_{1_i}\right|$

b) The mean-squared error loss (MSE) $= \frac{1}{n}\sum_{i=1}^{n}\left(estimated_T_{1_i} - MOLLI_T_{1_i}\right)^2$

where $n$ is the number of $T_1$ samples in a minibatch of the training dataset (batch size =80 $T_1$ maps). To select the optimal loss function and network size, the training dataset was further split into subsets of 80% training and 20% validation of the network for the optimization experiments only. The stochastic gradient descent optimizer with an initial learning rate of 1E-8 and momentum of 0.8 was used to train the network for 2000 epochs with a batch size of 80 maps. The same hyperparameters were used to train all models on native and post-contrast datasets, except for networks with $S = 3$ or $M_l = 3$, where a batch size of 60 $T_1$ maps was used. MyoMapNet took 8.3±2.5 hours to be trained using the native $T_1$ dataset.

## MyoMapNet Evaluation

The performance of MyoMapNet was evaluated using the independent testing dataset. The simulated testing dataset contained 16,000 samples, containing $T_1$ ranging from 400 to 1800ms. For in-vivo studies, we used native $T_1$ maps from 144 patients (432 MOLLI-5(3)3 $T_1$ maps) and post-contrast $T_1$ maps from 107 patients (324 MOLLI-4(1)3(1)2 $T_1$ maps). For each $T_1$ map, we calculated 3 sets of $T_1$ estimates: (a) *standard MOLLI* $T_1$ mapping using 3-parameter curve fitting from all collected images (8, or 9 images in native or post-contrast $T_1$ mapping, respectively), (b) *abbreviated MOLLI* using only 5 native (MOLLI-5) or 4 post-contrast MOLLI (MOLLI-4) images using standard 3-parameter curve fitting, and (c) *MyoMapNet* using only 5 native or 4 post-contrast $T_1$ weighted images, retrospectively extracted from the standard MOLLI dataset. The ECV values were calculated for patients who had both native and post-contrast $T_1$ mapping available in the testing dataset (75 patients).

## Data analysis

Endocardial and epicardial contours were drawn to measure global $T_1$ values. **Supplementary Figure S3** shows example contours for 4 different subjects. The mean and standard deviation of $T_1$ or ECV from the LV myocardium were calculated. The global $T_1$ was measured by averaging the myocardial $T_1$ from three different slices. Native, post-contrast $T_1$, and ECV values were reported as mean ± standard deviation. The 95% confidence interval (95% CI) of the mean was reported to represent a range of values that contains the true mean of the population with 95% certainty. The precision of $T_1$ maps was assessed using the standard deviation of $T_1$ values within the myocardium. Bland-Altman analysis was used to characterize per slice and per patient $T_1$ estimated by two different methods (abbreviated MOLLI vs. standard MOLLI, and MyoMapNet vs. standard MOLLI) and reported as the mean difference of $T_1$ estimations and 95% limits of agreements.

Group differences in per-slice $T_1$ measurements were assessed using linear mixed-effects models with compound symmetry covariance structure (41). In these models, the pair-wise group difference was fitted as the outcome, and the subject was included as a random intercept. The average and standard deviation of $T_1$ were compared between two methods using the paired Student's t-test. Bonferroni correction was utilized to account for three pair-wise group comparisons. All tests were two-sided and the nominal level of statistical significance was 0.0167.

# Results

## Network optimization

The native $T_1$ estimation errors between MyoMapNet and standard MOLLI-5(3)3 in the validation dataset at different loss functions, number of layers, and number of nodes per layer (18 networks in total) are summarized in **Table 1**. In general, MAE loss showed consistently lower errors than MSE at different network sizes and was therefore used to train all MyoMapNet models. For MAE, small changes in $T_1$ estimation errors were reported between MyoMapNet and MOLLI-5(3)3 at different network sizes with slightly lower STD for bigger networks. MyoMapNet of 5 hidden layers ($M_l = 2$) and 400, 400, 200, 200, and 100 nodes ($S = 2$) at each layer with a total of 305,401 trainable parameters, was therefore used for all $T_1$ estimation models to balance between small estimation errors and lower chances of overfitting due to increased number of parameters.

## Numerical Simulations

**Figure 3** shows the differences in estimated $T_1$ between the MOLLI-5 (left) and MyoMapNet (right) vs MOLLI-5(3)3, simulated at SNR levels of 100, 40, and 20. For higher SNR, the estimated $T_1$ values using MOLLI-5 exhibit a systematic error as a function of $T_1$ values. For the same SNR, MyoMapNet shows no systematic error for different $T_1$ values. As SNR decreases, there is an increase in the error of estimated $T_1$ values using both techniques, but with lower error in MyoMapNet (0±15 ms, 95% CI [-0.3, 0.1]) compared to MOLLI-5 (-3±32 ms, and 95% CI [-4, -3]; P <0.001) relative to MOLLI-5(3)3 at SNR=40. At SNR=20, the mean difference between MyoMapNet and MOLLI-5(3)3 was (0±31ms, 95% CI [-1, 1]) compared to MOLLI-5 (-3±54 ms, 95% CI [-4, -3]; P <0.001).

**In-vivo Studies:**

**Figure 4** shows example native $T_1$ maps reconstructed with the MOLLI-5(3)3, MOLLI-5, and MyoMapNet in four subjects. The $T_1$ maps using MyoMapNet showed sharp myocardial $T_1$ boundaries with more homogeneous $T_1$ estimations similar to that of MOLLI-5(3)3 maps when compared to the MOLLI-5 method. This inhomogeneity in MOLLI-5 $T_1$ estimations is more apparent in the blood pool, affecting the accuracy and precisions of the estimated $T_1$ values.

The mean $T_1$ values and precision for myocardial native and post-contrast $T_1$, and ECV maps are summarized in **Table 2**. MyoMapNet yields $T_1$ and ECV values similar to those of the standard MOLLI methods with lower estimation errors compared to abbreviated MOLLI methods (1230±63, 1200±45, and 1199±46 ms for native $T_1$ values; 548±63, 563±45, and 565±47 ms for post-contrast $T_1$, and 29.3±5.6, 27.3±3.5, and 27.1±4% for ECV values using abbreviated MOLLI, MyoMapNet, and standard MOLLI methods, respectively). For native $T_1$ precision, the average precision of MyoMapNet $T_1$ estimations (120 ms, 95% CI [117, 122]) was superior to MOLLI-5 (148 ms, 95% CI [144, 151]; P <0.01) and inferior to standard MOLLI-5(3)3 (112 ms, 95% CI [109, 114]; P <0.01). In post-contrast $T_1$, the average precision of MyoMapNet $T_1$ estimations (44 ms, 95% CI [42, 45]) was similar to the standard MOLLI-4(1)3(1)2 (43 ms, 95% CI [41, 44]; P=0.05) and significantly higher than MOLLI-4 (52 ms, 95% CI [50, 54]; P <0.01).

Bland-Altman plots for $T_1$, post-contrast $T_1$, and ECV for three different approaches (**Figure 5)** show similar measurements for MyoMapNet when compared to MOLLI, with increased measurement error for abbreviated MOLLI. Similar results were observed for per slice analysis of the three methods (**Supplementary Figure S4**).

The duration for estimating $T_1$ maps was 24.2±0.5 sec by a 3-parameter fitting model without motion correction using Matlab on CPU, and 2±0.01 ms by MyoMapNet using Python on GPU.

## Discussion

In this study, we evaluated the potential of a fully connected neural network for native and post-contrast myocardial $T_1$ and ECV mapping from 4-5 $T_1$ weighted images, reducing the scan time of myocardial $T_1$ mapping by 50% without compromising accuracy or precision.

Respiratory motion in myocardial tissue mapping causes artifacts that adversely impact the quantification of tissue relaxation times. To minimize respiratory motion artifacts, imaging during breath-holding or use of respiratory slice tracking have been used (6,42–48). For MOLLI, images for each slice are collected in a single breath-hold. However, even with breath-holding, patients often have respiratory motion-induced drift, which creates artifacts in maps. Therefore, respiratory motion correction is still being recommended (1). Similarly, for free-breathing tissue characterization, while slice tracking reduced through-plane motion, motion correction is necessary to align images (44,46). Furthermore, motion correction will improve the reproducibility and robustness of tissue mapping (49). One of the advantages of MyoMapNet is reducing the sensitivity to respiratory motion. **Supplementary Figure S5** shows example images demonstrating the presence of motion in standard MOLLI data, despite breath-holding. In this study, we used retrospectively collected data and were not able to perform a head-to-head comparison of the impact of shorter scan time on the motion. Further studies are warranted to further investigate the potential advantages of MyoMapNet in reducing motion sensitivity.

The numerical simulations of native $T_1$ MOLLI with SNR=100 showed systematic errors as a function of the $T_1$ value increments associated with MOLLI-5 relative to the standard MOLLI-5(3)3 estimations. The MyoMapNet learns a correction function to minimize such systematic errors and create unbiased $T_1$ estimations at different ranges of $T_1$ values. As more noise was added to the $T_1$-weighted signals, the MOLLI-5 exhibited higher $T_1$ estimation errors. MyoMapNet significantly reduces the errors associated with noise, while correcting for systematic errors. Similar observations were seen in the in-vivo data for native and post-contrast $T_1$ mapping, where MyoMapNet showed significantly lower $T_1$ and ECV estimation errors compared to the abbreviated MOLLI methods.

Machine learning is rapidly improving the workflow of myocardial tissue characterization in CMR (50). Recent studies have demonstrated the potential for deep learning in automating analysis workflow and quality control (39,51,52). These methods could automatically perform motion correction, segmentation, and parameter quantification, thereby reducing the burden and observer-related variability of manual analysis. The MyoMapNet could be easily integrated with an automated analysis and quality control method to simplify the workflow.

In this study, we reduced the number of $T_1$-weighted images to 4-5 images for estimating $T_1$ values without rigorously studying the impact of the number of $T_1$-weighted images. Further studies are warranted to investigate the optimal number of $T_1$-weighted images for myocardial $T_1$ mapping. For MOLLI, after LL, samplings along the relaxation curve are separated by cardiac cycle. The choice of inversion time can potentially impact the accuracy and precision of tissue mapping (53). Hence, the effective LL times are determined by the RR interval length. In a single LL experiment, only the first $T_1$-weighted image has a short LL time (<RR interval length), while the rest of the images have a long LL time and hence are less sensitive to $T_1$. This is more prominent in post-contrast $T_1$ mapping where $T_1$ times range from 100 to 600 ms, which will impact the quality of the $T_1$ map. Other acquisition schemes, such as 4(1)1 for native $T_1$, or 3(1)1 or 2(1)2 for post-contrast $T_1$, could add another LL anchor at the beginning of the relaxation curve to increase the susceptibility of $T_1$, without increasing the scan time more, which necessitates further investigation.

The fully-connected neural network architecture adopted in MyoMapNet was more convenient for the pixel-wise, short, fixed-length input data (10 in native or 8 in post-contrast $T_1$ MOLLI) than convolutional-based neural networks to learn features from all combinations of the input data with a maximum receptive field. Similar fully-connected based architectures were utilized for simultaneous calculation of $T_1$ and $T_2$ maps from MR fingerprinting acquisitions in cardiac and brain imaging (37,54). On the other hand, a one-dimensional convolutional neural network was used to estimate $T_1$ and $T_2$ maps from a relatively longer input of 220 signal intensity values and inversion time stamps of BLESSPC acquisitions (36). Other architectures for MyoMapNet can be investigated for $T_1$ estimations in future studies. MyoMapNet was also successful in estimating $T_1$ values in two MOLLI sequences, MOLLI-5(3)3, and MOLLI-4(1)3(1)2, which indicates its flexibility and potential generalizability to other $T_1$ and $T_2$ sequences. Future studies are warranted to investigate the generalizability of MyoMapNet to alternative $T_1$ and $T_2$ sequences.

Our study has several limitations. MyoMapNet was validated using retrospectively acquired data derived from standard MOLLI-5(3)3 and MOLLI-4(1)3(1)2, instead of collecting dedicated data with 4-5 $T_1$-weighted images. We only evaluated MyoMapNet using MOLLI data, a similar concept can be investigated for other $T_1$ mapping sequences. We did not perform any motion correction in $T_1$ mapping. We used MOLLI sequence for training, and it is widely known that MOLLI has intrinsic under-estimation. Further studies should be pursued to investigate the potential of MyoMapNet to improve measurement accuracy. We used a large dataset of patients with different clinical indications, therefore the clinical utility of MyoMapNet was not studied. Finally, data from a single vendor and field strength was used for training and the generalizability of the trained network for other field strengths or MR vendor should be studied.

## Conclusion

The MyoMapNet enables fast and precise myocardial $T_1$ mapping quantification from only 4-5 $T_1$-weighted images acquired after a single look-locker sequence, leading to shorter scan time and rapid map reconstruction.

## Funding Information

Reza Nezafat receives grant funding by the National Institutes of Health (NIH) 1R01HL129185, 1R01HL129157, 1R01HL127015 and 1R01HL154744 (Bethesda, MD, USA); and the American Heart Association (AHA) 15EIA22710040 (Waltham, MA, USA). All other authors have reported that they have no relationships relevant to the contents of this paper to disclose.

## Data Availability Statement

The codes of MyoMapNet are openly available in Harvard dataverse at (https://dataverse.harvard.edu/dataverse/cardiacmr), reference number (https://doi.org/10.7910/DVN/VYV31B).

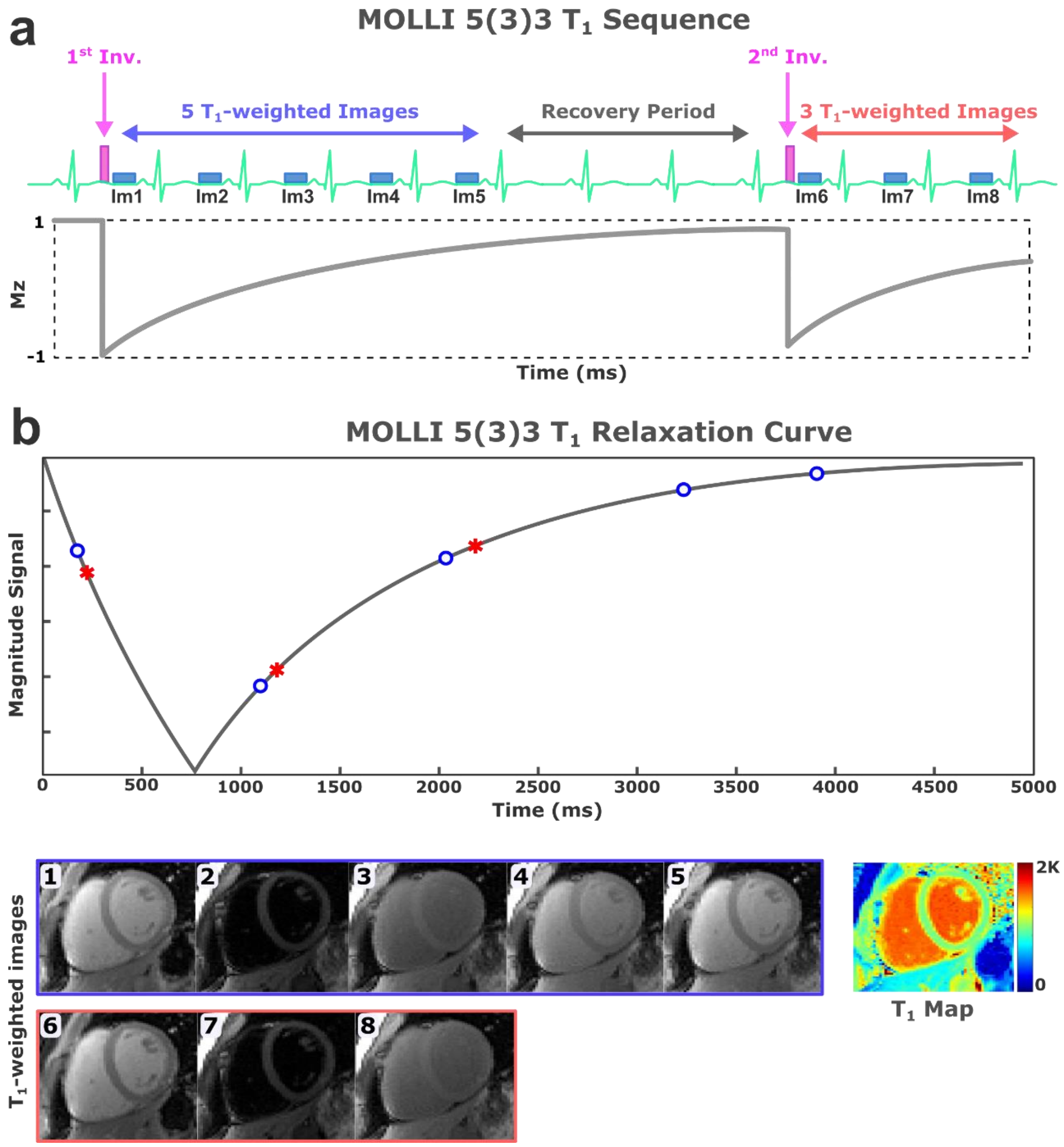

**Figure 1.** MOLLI-5(3)3 sequence (a) and associated signal recovery (b): two look-locker inversion pulses are performed to acquire eight $T_1$-weighted images with a recovery period of 3 heartbeats between look-lockers. In MyoMapNet, images collected after the first inversion pulse are used for estimating the $T_1$ values.

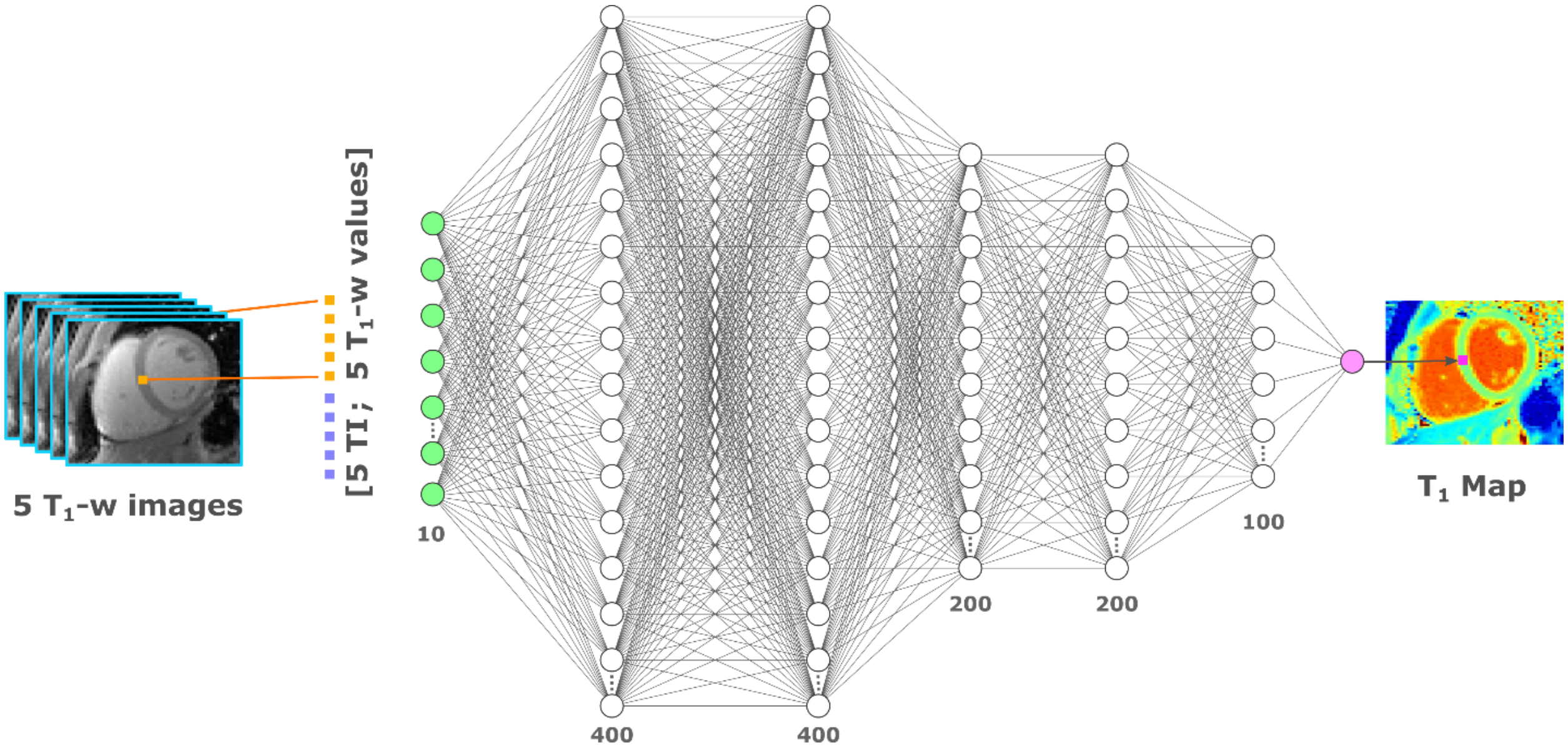

**Figure 2.** MyoMapNet architecture: MyoMapNet uses a fully-connected neural network for estimating pixel-wise $T_1$ values from $T_1$-weighted images collected after a single look-locker inversion pulse (i.e. the first 5 images of MOLLI-5(3)3 or first 4 images of MOLLI-4(1)3(1)2). For each voxel, the signal values from 5 $T_1$-weighted images are concatenated with their corresponding look-locker times and used as the network input (i.e. 10×1) for native $T_1$ mapping. The input values are fed to a fully-connected network with 5 hidden layers with 400, 400, 200, 200, and 100 nodes each layer, respectively. The output is the estimated $T_1$ value at each voxel.

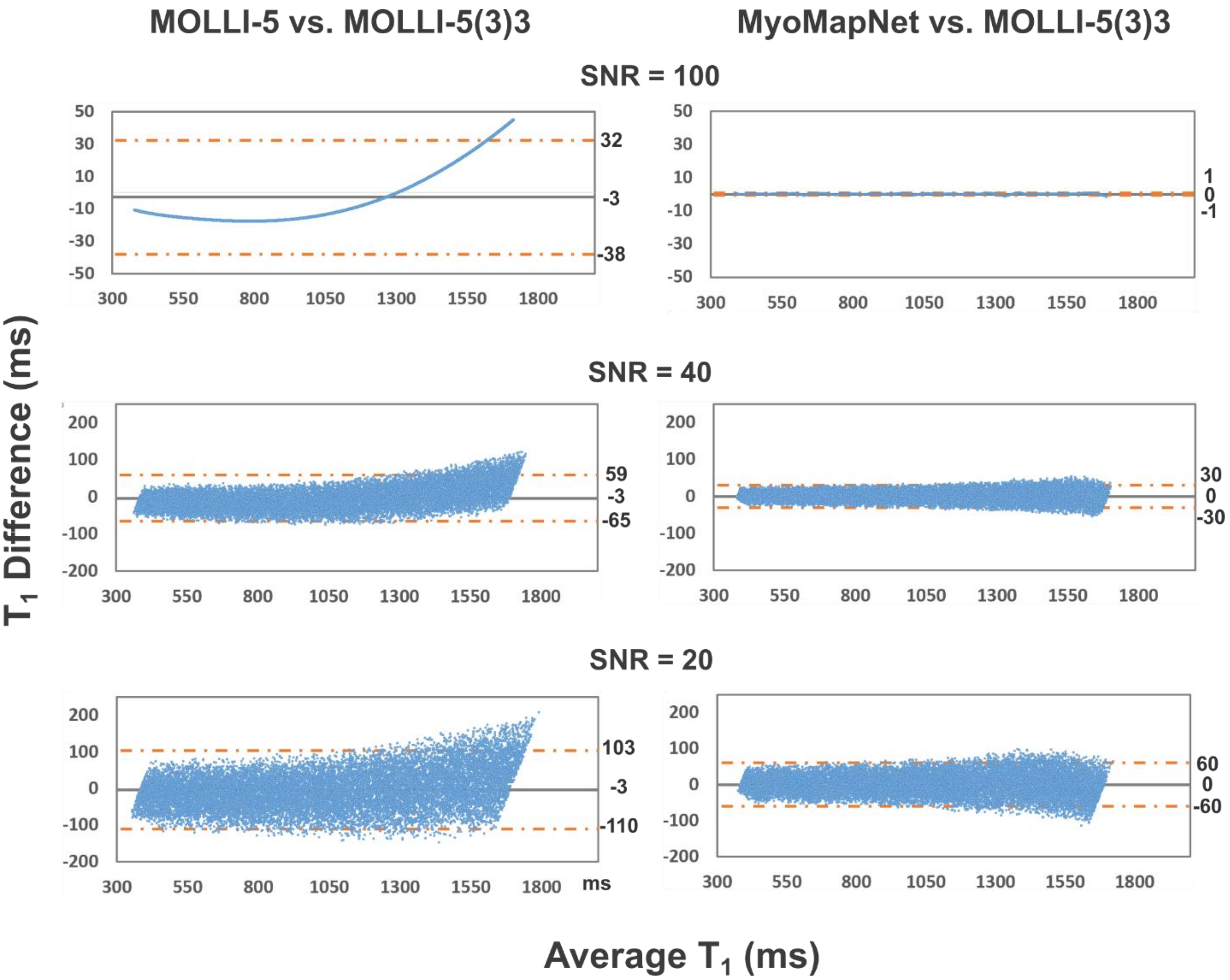

**Figure 3.** Bland-Altman plots showing the mean difference and 95% limits of agreement between the simulated $T_1$ values by the MOLLI-5 vs. MOLLI-5(3)3 (left column) and MyoMapNet vs. MOLLI-5(3)3 (right column) for SNR of 100, 40, and 20. MOLLI-5 yields a systematic error in $T_1$ estimations, which is corrected in MyoMapNet.

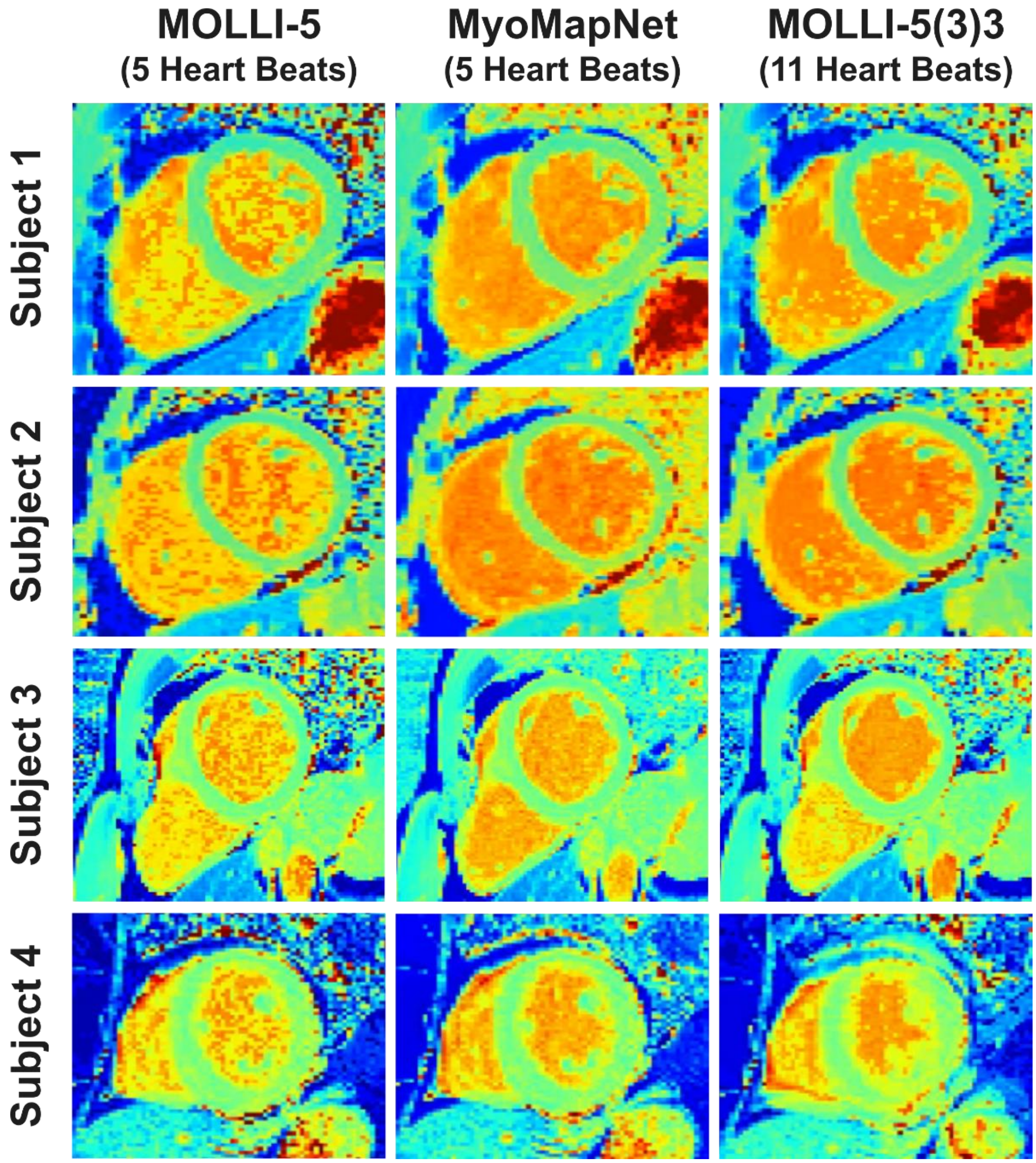

**Figure 4.** Native $T_1$ maps from four patients, reconstructed using MOLLI-5 (using only 5 $T_1$ weighted images with 3-parameter fitting), MyoMapNet, and MOLLI-5(3)3 with a 3-parameter fitting model. MoyMapNet yield maps with more homogenous signal compared to MOLLI-5.

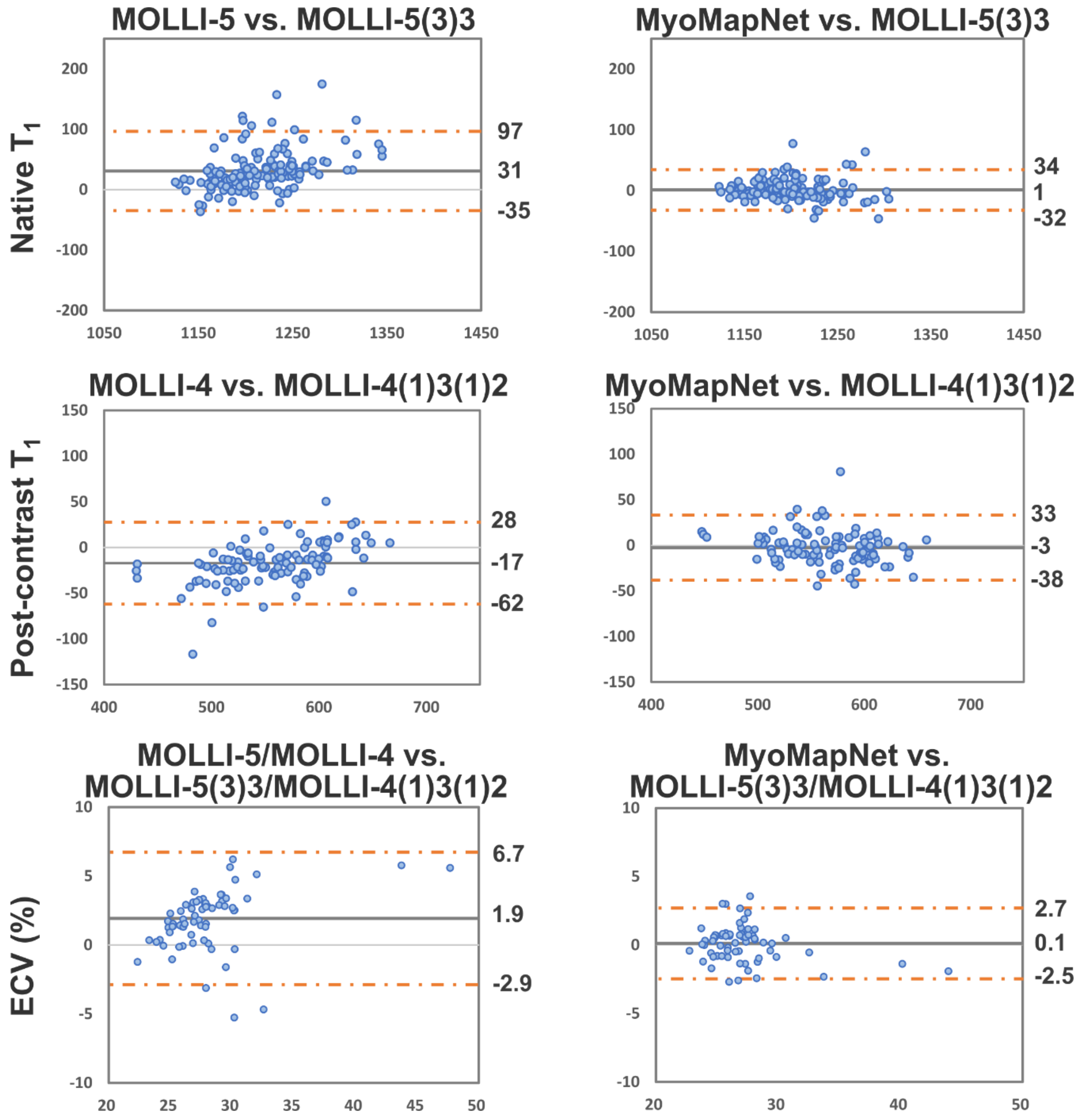

**Figure 5.** Bland-Altman plots showing the mean difference and 95% limits of agreement for global native, post-contrast $T_1$, and ECV for the abbreviated MOLLI-5/MOLLI-4 vs. standard MOLLI-5(3)3/MOLLI-4(1)3(1)2 (left column) and MyoMapNet vs. standard MOLLI-5(3)3/MOLLI-4(1)3(1)2 (right column). Each dot represents the values calculated from one patient.

**Table 1.** Mean difference (±STD) in $T_1$ values between MyoMapNet and MOLLI-5(3)3 at different loss functions (MSE, and MAE), network depth (3, 5, and 7 layers) and network size (S=1, 2, and 3).

| | **MSE Loss** | | | **MAE Loss** | | |
|---|---|---|---|---|---|---|
| | **3 layers** | **5 layers** | **7 layers** | **3 layers** | **5 layers** | **7 layers** |
| **S=1** | 0±25 | 12±30 | 9±24 | 3±27 | 5±24 | 5±24 |
| **S=2** | 12±23 | 5±24 | 5±22 | 6±24 | 3±24 | 4±22 |
| **S=3** | 11±23 | 18±21 | -2±24 | 3±24 | 3±23 | 3±22 |

**Table 2.** $T_1$/ECV values and precision (standard deviation of measurements within myocardium) for the MOLLI-5, MyoMapNet, and MOLLI-5(3)3 in the testing dataset of 144, 107 and 75 patients for native $T_1$, post-contrast, and ECV, respectively.

| | Myocardial $T_1$ and ECV | | | Precision | | |
|---|---|---|---|---|---|---|
| | **MOLLI-5/ MOLLI-4** | **MyoMapNet** | **MOLLI-5(3)3/ MOLLI-4(1)3(1)2** | **MOLLI-5/ MOLLI-4** | **MyoMapNet** | **MOLLI-5(3)3/ MOLLI-4(1)3(1)2** |
| **Native $T_1$ (ms)** | $1230 \pm 63^{*\S}$ | $1200 \pm 45$ | $1199 \pm 46$ | $148 \pm 34^{*\S}$ | $120 \pm 27^{*}$ | $112 \pm 27$ |
| **Post-contrast $T_1$ (ms)** | $548 \pm 63^{*\S}$ | $563 \pm 45$ | $565 \pm 47$ | $52 \pm 11^{*\S}$ | $44 \pm 11$ | $43 \pm 10$ |
| **ECV (%)** | $29 \pm 6^{*\S}$ | $27 \pm 4$ | $27 \pm 4$ | - | - | - |

$^{*}$p-value < 0.01 when compared to MOLLI.
$^{\S}$p-value < 0.01 when compared to MyoMapNet.

**Table S1.** Number of parameters at different MyoMapNet depth (3, 5, and 7 layers) and network size (S=1, 2, and 3).

| | 3 layers | 5 layers | 7 layers |
|---|---|---|---|
| **S=1** | 27401 | 77701 | 128001 |
| **S=2** | 104801 | 305401 | 506001 |
| **S=3** | 232201 | 683101 | 1134001 |

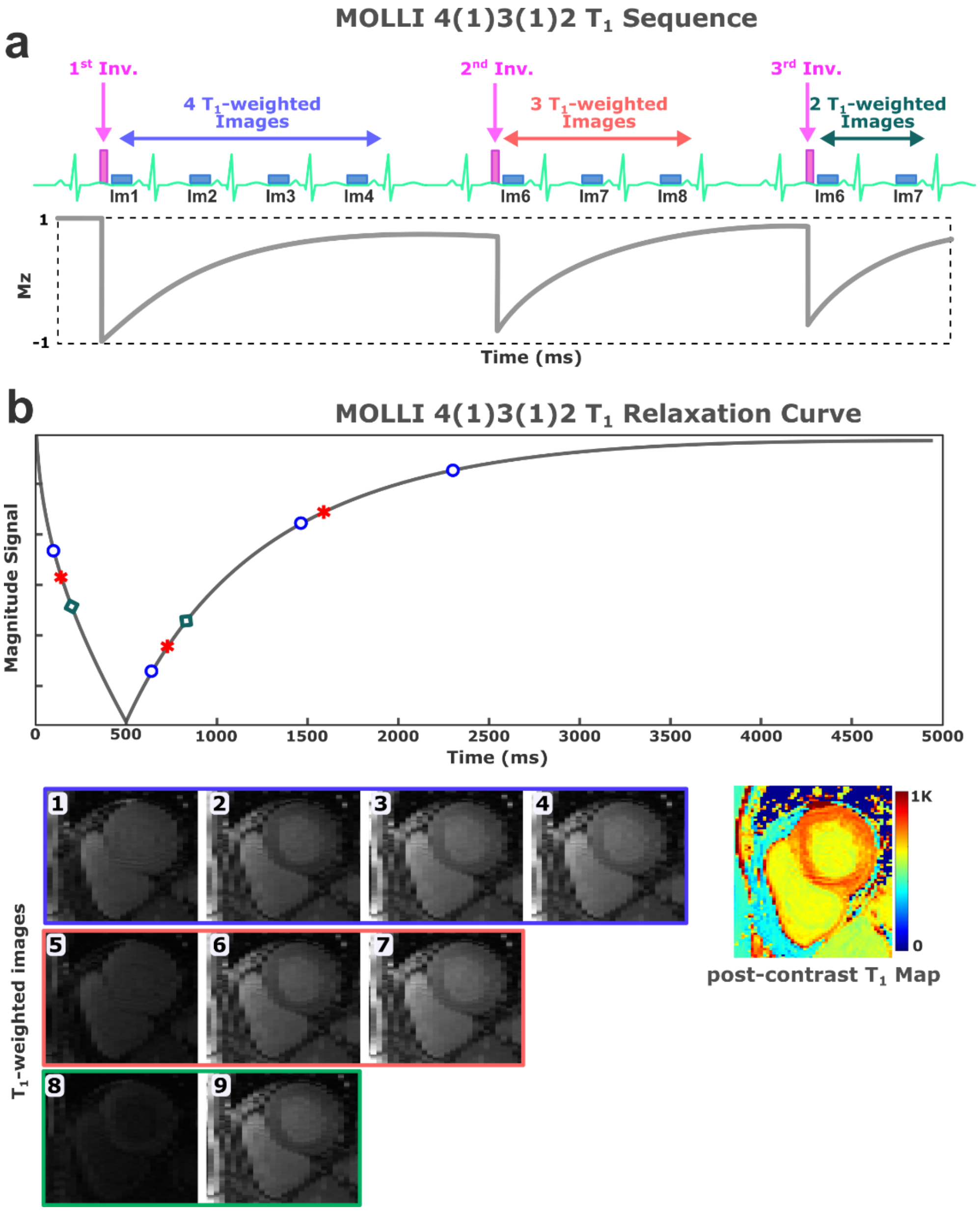

**Figure S1.** MOLLI-4(1)3(1)2 sequence (a) and associated signal recovery (b): three look-locker inversion pulses are performed to acquire nine $T_1$-weighted images with a recovery period of one heartbeat between look-lockers. In MyoMapNet, images collected after the first look-locker are used for estimating the $T_1$ values.

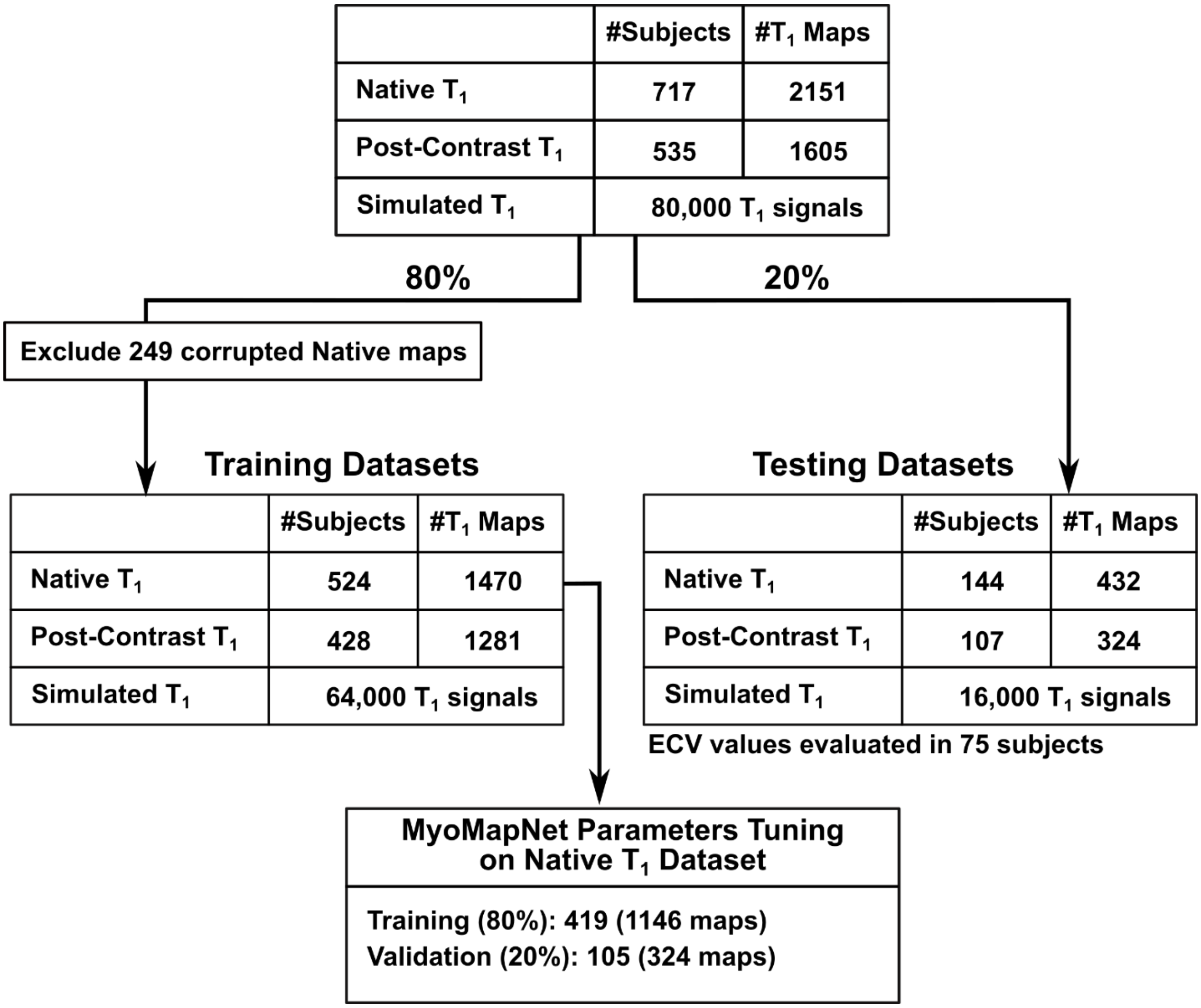

**Figure S2.** Flowchart for MyoMapNet training, testing, and parameter-tuning validation splits in the native, post-contrast, and simulated $T_1$ datasets. ECV values were evaluated in 75 subjects who had both native and post-contrast $T_1$ data available in the testing dataset.

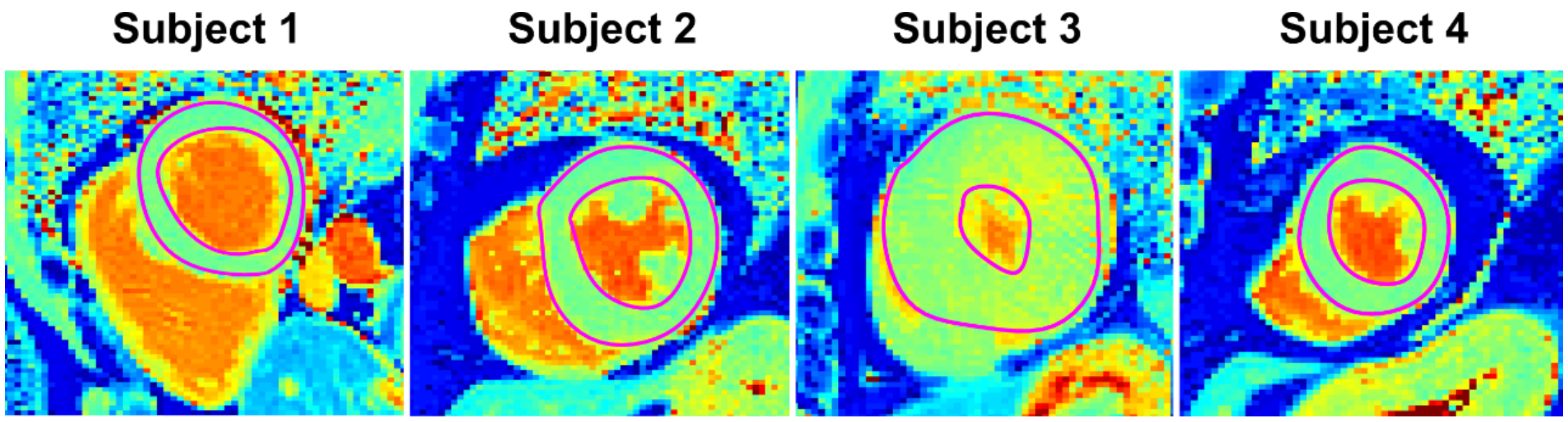

**Figure S3.** Example native $T_1$ maps with left ventricular myocardial segmentation used for analyzing $T_1$ estimations by MyoMapNet, MOLLI-5, and standard MOLLI-5(3)3.

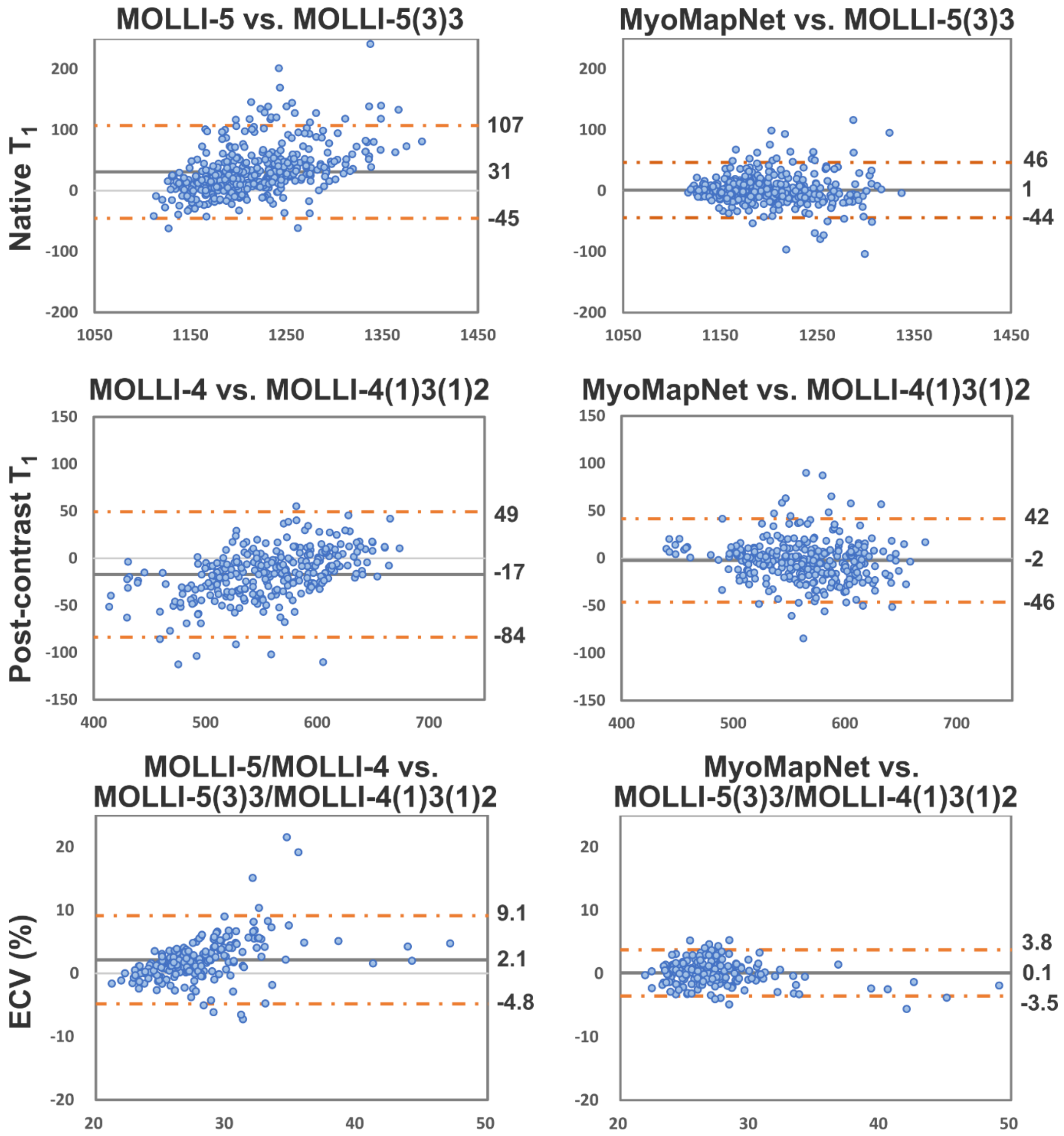

**Figure S4.** Bland-Altman plots showing per slice analysis of the mean difference and 95% limits of agreement for native, post-contrast $T_1$, and ECV for the abbreviated MOLLI-5/MOLLI-4 vs. standard MOLLI-5(3)3/MOLLI-4(1)3(1)2 (left column) and MyoMapNet vs. standard MOLLI-5(3)3/MOLLI-4(1)3(1)2 (right column). Each dot represents the values calculated from one slice.

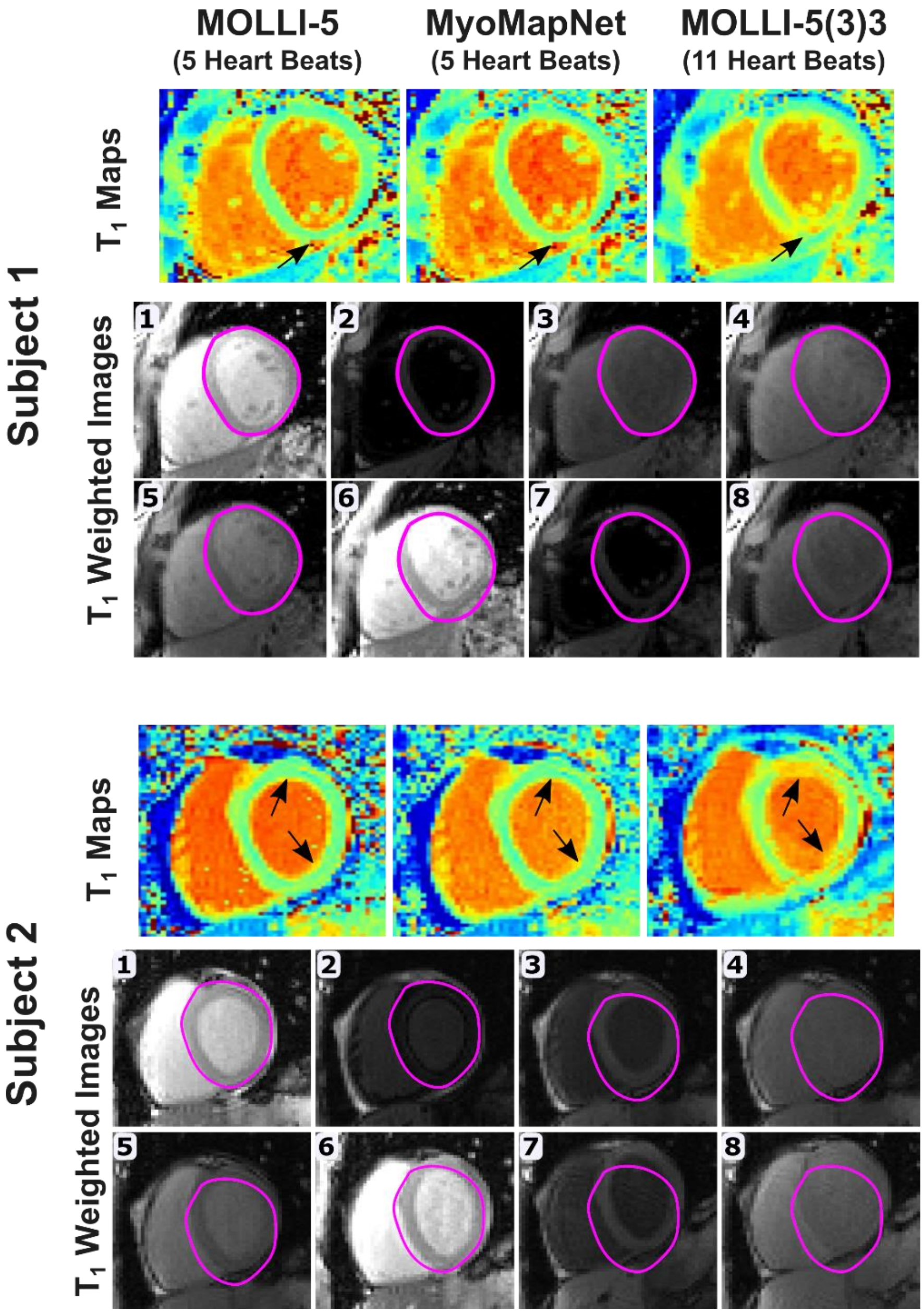

**Figure S5.** Myocardial $T_1$ maps calculated using MOLLI-5 with a 3-parameter fitting model, MyoMapNet, and MOLLI-5(3)3, demonstrating the impact of respiratory motion (black arrows) in two subjects. The epicardial delineation of image #1 was copied to all other $T_1$-weighted images to show the degree of motion between images. Both subjects were instructed to hold their breath during the scan.